\documentstyle {article}

\newcommand{\Section}[1]{\section{#1}\setcounter{equation}{0}}

\def\p{\partial}
\def\be{\begin{equation}}
\def\bea{\begin{eqnarray}}

\def\nn{\nonumber}
\def\ll{\lambda}
\def\l{\label}
\def\D{\Delta}
\def\ee{\end{equation}}
\def\eea{\end{eqnarray}}
\def\o{\over}
\begin{document}
\begin{titlepage}
\begin{center}
\vskip .2in
\hfill
\vbox{
    \halign{#\hfil         \cr
           hep-th/9704082 \cr
           } 
      }  
\vskip 1cm
{\large \bf Global Conformal Invariance in D-dimensions}\\
{\large \bf and}\\
{\large \bf Logarithmic Correlation Functions}
\vskip .6in
{\bf A. M. Ghezelbash$^{*,+,}$} \footnote {e-mail address:
amasoud@physics.ipm.ac.ir}\\
{\bf V. Karimipour$^{*,\dagger,}$} \footnote{e-mail address:
vahid@physics.ipm.ac.ir}\\
\vskip .25in
{\em
$^*$Institute for Studies in Theoretical Physics and Mathematics,
P.O. Box 19395-5531, Tehran, Iran.\\
$^+$Department of Physics, Alzahra University, Vanak,
Tehran 19834, Iran.\\
$^\dagger$Department of Physics, Sharif University of Technology,
P. O. Box 19365-9161, Tehran, Iran.}
\vskip 1cm
\end{center}
\begin{abstract}
  We define transformation of multiplets of fields (Jordan cells) under the
D-dimensional conformal group, and calculate two and three point functions of
fields, which show logarithmic behaviour. We also show how by a formal
differentiation procedure, one can obtain n-point function of logarithmic field
theory from those of ordinary conformal field theory.
  \end{abstract}
\end{titlepage}
\newpage
\Section{Introduction}

Recently, Logarithmic Conformal Field Theories (LCFT) have been studied in a
series of papers \cite{1}-\cite{15}, both for their pure theoretical interest
concerning
the structure and classification of conformal field theories \cite{1,7,8,9,12}
and
for their relevance in some physical systems \cite{2,3,4,5,6,10} which are
described by
non-unitary or non-minimal conformal models.\newline
All such works have dealt with two dimensional conformal field theory \cite{1}
relying heavily on the underlying Virasoro algebra, and have described how
the appearance of logarithmic singularities is related to the modification
of the representation of the Virasoro algebra.\newline
In this paper we will try to understand LCFT's from yet another point of view,
that's we consider d-dimensional conformal invariance.\newline
Although our results are based on a generalization of the basic idea of \cite{1}
, we hope that by changing or simplifying the context of study, we can add a
little bit to the understanding of the subject in general.\newline
As is well known, one of the basic assumptions of conformal field theory is
the existence of a family of operators, called scaling fields, which transform
under scaling $S: x\rightarrow x'=\ll x$, simply as follows:
\be  \l{scalet}
\phi (x) \rightarrow \phi '(x')=\ll ^{-\Delta} \phi (x),
\ee
where $\Delta$ is the scaling weight of $\phi (x)$.
It is also assumed that under the conformal group, such fields transform as,
\be \l{conft}
\phi (x) \rightarrow \phi '(x')=\parallel {{\p x'}\o {\p x}}\parallel ^{
{{-\Delta} \o {d}}}
\phi (x),
\ee
where $d$ is the dimension of space and
$\parallel {{\p x'}\o{\p x}}\parallel $ is the Jacobian
of the transformation. Equation (\ref{conft}) which encompasses eq.
(\ref{scalet})
defines the transformation of the quasi-primary fields. For future use we
note that the Jacobian equals $\ll ^d$ for scaling transformation and
$\parallel x\parallel ^{-2d}$ for the Inversion transformation $I: x\rightarrow
 x'={{x}\o
{{\parallel x\parallel}^2}}$, being unity for the other elements of the
conformal group.
Combination of (\ref{conft}) with the definition of symmetry of the correlation
functions, i.e.:
\be  \l{invarcor}
<\phi '_1(x'_1)\cdots\phi '_N(x'_N)>=
<\phi _1(x'_1)\cdots\phi _N(x'_N)>,
\ee
allows one to determine the two and the three point functions up to a
constant and the four point function up to a function of the cross ratio.
\newline
{\it It's precisely the assumption that scaling fields constitute irreducible
representations
of the scaling transformation, which imposes power law singularity on the
correlation functions}. As we will see, relaxing this assumption, one naturally
 arrives
at logarithmic singularities. It also leads to many other pecularities,
in the relation between correlation functions.
To begin with, we consider a multiplet of fields,
\be\Phi =\left (\begin{array}{c} \phi _1\\ \phi _2 \\ \vdots \\ \phi _n
\end{array} \right ),\ee and note that  under scaling $x\rightarrow \ll x$, the
most general form of the transformation of $\Phi$ is,
\be \l{mt}
\Phi (x)\rightarrow \Phi '(x')=\ll ^{T'}\Phi (x)
\ee
where $T'$ is an arbitrary matrix. More generally, we replace (\ref{mt}) by,
\be \l{confgt}
\Phi (x) \rightarrow \Phi '(x')=\parallel {{\p x'}\o {\p x}}\parallel ^T
\Phi (x).
\ee
where $T$ is an $n\times n$ arbitrary matrix. When $T$ is diagonalizable, one
arrives at ordinary scaling fields by redefining $\Phi$, so that all the fields
transform as $1$-dimensional representation. Otherwise, following \cite{13} we
assume that $T$ has Jordan form,
\be  \l{t}
T=\pmatrix { {{-\Delta}\o {d}}&0&\cdots&0\cr
1&{{-\Delta}\o {d}}&\cdots&0\cr
0&1&\ddots&0\cr
0&\cdots&1&{{-\Delta}\o {d}} }.
\ee
Rewriting $T$ as ${{-\Delta}\o{d}}1+J$, where $J_{ij}=\delta _{i-1,j}$,
eq. (\ref{confgt}) can be written in the form,
\be \l{confgtlam}
\Phi (x) \rightarrow \Phi '(x')=\parallel {{\p x'}\o {\p x}}\parallel ^{{{-
\Delta} \o {d}}}\Lambda _x
\Phi (x),
\ee
where $\Lambda _x=\parallel {{\p x'}\o {\p x}}\parallel ^{J}$ is a lower
triangular matrix of the form,
\be \l{lambda}
{(\Lambda _x)}_{ij}={{{\{\ln \parallel {{\p x'}\o{\p x}}\parallel\}}^{i-j}}\o
{(i-j)!}},\qquad  {(\Lambda _x)}_{ii}=1,
\ee
i. e. for $N=2$ we have,
\bea \l{N2}
\phi _1'(x')&=&{\parallel {{\p x'}\o {\p x}}\parallel}^{{{-\Delta} \o {d}}}
\phi _1(x),\nn\\
\phi _2'(x')&=&{\parallel {{\p x'}\o {\p x}}\parallel}^{{{-\Delta} \o {d}}}
\big( \ln \parallel {{\p x'}\o {\p x}}\parallel \phi _1(x)+\phi _2(x)
\big).
\eea
An important point is that the top field $\phi _1(x)$ {\it always} transform as an ordinary quasi-primary
field. A most curious propety of the transformation (\ref{confgtlam}) is
that each field $\phi _{k+1}$ transforms as if it is a formal derivative of
$\phi _k$ with respect to
${{-\Delta} \o {d}}$,
\be  \l{der}
\phi _{k+1}(x)={1 \o k}{{\p}\o {\p ({{-\Delta }\o {d}})}}\phi _k(x).
\ee
This formal relation which determines the transformation of all the fields
of a Jordan cell from that of the top field $\phi _1$, essentially means that
with due care, one can determine the correlation functions of the lower fields
from those of the ordinary top fields simply by formal differentiation.
We will elucidate this point later on. The phrase with due care in the previous
statement refers to the two point function of fields within {\it the same}
Jordan cell.
\Section{The Two Point Functions}

We already know by standard arguments \cite{15} that the two point function of
the top
fields $\phi _\alpha$ and $\phi _\beta$ belonging to two different Jordan cells
$(\Delta _\alpha ,n)$ and $(\Delta _\beta ,m)$ vanishes, i.e.:
\be \l{zero}
<\phi _\alpha (x)\phi _\beta (y)>={{A {\delta _{{\Delta _\alpha},{\Delta
_\beta}}}}\o{\parallel x-y\parallel^{2\Delta _\alpha}}}.
\ee
Due to the observation
(\ref {der}), it follows that the two point function of all the fields of two
different Jordan cells with respect to each other vanish. Therefore
in this section we calculate the two point function of the fields within the
same
Jordan cell. As we will see logarithmic conformal symmetry gives many
interesting
and unexcepted results in this case. Let's denote the matrix of two point
functions  $<\phi _i (x)\phi _j (y)>$ for all $\phi _i,\phi _j \in (\Delta ,n)$
by $G(x,y)$, then from rotation and translation symmetries, this matrix should
depends only on $\parallel x-y \parallel$. From scaling symmetry and using
(\ref{conft}) and (\ref{invarcor}),
we will have,
\be \l{ss}
\Lambda G(\parallel x-y \parallel )\Lambda ^t=\ll ^{2\D} G(\ll \parallel x-y
\parallel ),
\ee
where $\Lambda =\ll ^{dJ}$, and from inversion symmetry, we have,
\be \l{is}
\Lambda _xG(\parallel x-y \parallel )\Lambda _y^t=\parallel x-y\parallel
^{-2\D _\alpha}
G({{\parallel x-y\parallel}\o{\parallel x\parallel \parallel y\parallel}}),
\ee
where \bea \l{lamx}\Lambda _x &=&\parallel x\parallel ^{-2dJ},\nn\\
\Lambda _y &=&\parallel y\parallel ^{-2dJ}.\eea
Defining the matrix $F$ as $G(\parallel x-y\parallel )={{F(\parallel x-y
\parallel )}\o {
{\parallel x-y \parallel }^{2\D}}}$, we will have from (\ref {ss}) and
(\ref {is}),
\be \l{ssf}
\Lambda F(\parallel x-y \parallel )\Lambda ^t=F(\ll \parallel x-y \parallel ),
\ee
and
\be \l{isf}
\Lambda _xF(\parallel x-y \parallel )\Lambda _y^t=
F({{\parallel x-y\parallel}\o{\parallel x\parallel \parallel y\parallel}}).
\ee
For every arbitrary $\ll$, we now choose the points $x$ and $y$ such that
$\parallel x\parallel =\ll ^{{{-1}\o{4}}}$ and $\parallel y\parallel =\ll
^{{{-3}\o{4}}}$.
{\it It should be noted that in this way by varying $\ll$, we can span all
the points of
space}. From (\ref {lamx}), we will have $\Lambda _x=\Lambda ^{{1\o 2}}$
and $\Lambda _y=\Lambda ^{{3\o 2}}$, therefore eq. (\ref {ssf}) turns into,
\be \l{isff}
\Lambda ^{{1 \o 2}}F(\parallel x-y \parallel ){\big ( \Lambda ^{{3\o 2}} \big )
}^t=
F(\ll \parallel x-y\parallel).\ee Combining (\ref {ssf}) and (\ref {isff}) and
using
invertibility of $\Lambda$, we arrive at,
\be \l{f}
F=\Lambda ^{{1 \o 2}}F{\big (\Lambda ^t\big )}^{{-1}\o 2},
\ee
by iterating (\ref{f}), we will have $F=\Lambda F\Lambda ^t$, and by
rearranging, we will have,
\be \l{ff}
F\Lambda ^t=\Lambda F.
\ee
Expanding $\Lambda $ in terms of power of $\ln \ll$ as
$\Lambda =1+(d\ln \ll )J+{{(d\ln \ll )^2}\o {2!}}J^2+\cdots $ and comparing
both sides, we arrive at
\be \l{fff}
F{\big (J^t\big )}^k={\big ( J\big )}^kF,\qquad k=1,2,\cdots ,n-1.
\ee
Since ${\big ( J^k\big )}_{ij}=\delta _{i,j+k}$, we will have from (\ref {fff}),
\be \l{ffff}
F_{i,j-k}=F_{i-k,j},
\ee
which means that on each opposite diagonal of the matrix $F$, all the
correlations are equal. Moreover from $FJ=JF$, one obtains,
\be \l{f5}
\sum _{l=1}^n F_{il}\delta _{j,l+1}=\sum _{l=1}^n\delta _{i,l+1}F_{lj},
\ee
which means that if $j=1$ and $1<i\leq n$, then $F_{i-1,j}=0$, or
\be \l{finalf} F_{ij}=0 \qquad {\rm for}\qquad j=1 \qquad{\rm and}\qquad
1\leq i\leq n-1.\ee
Combining
this with (\ref {ffff}), we find that all the correlations above the opposite
diagonal are zero. In order to find the final form of $F$, we use eq.
(\ref {ssf})
again, this time in infinitesimal form, let $\Lambda =1+\alpha J+o(\alpha ^2)$
where $\alpha =d\ln \ll $, then from $\Lambda F(x)\Lambda ^t=F(\ll x)$, we have,
\be
\l{infi}
d\big (JF+FJ^t\big )=x{{dF}\o{dx}}.
\ee
Due to the property (\ref {ffff}) only the last column of $F$ should be found,
therefore
from (\ref {infi}) we obtain,
\bea \l{lc}
x{{dF_{1,n}}\o {dx}}&=&dF_{1,n-1}\equiv 0,\nn\\
x{{dF_{i,n}}\o {dx}}&=&2dF_{i,n-1},\qquad {\rm if}\quad i>1
\eea
which upon introducing the new variable $y=2d\ln x$ gives,
\be \l{fs}
F_{1,n}=c_1,\quad
F_{2,n}=c_1y+c_2,\quad
F_{3,n}={ 1\o 2}c_1y^2+c_2y+c_3,\quad
{\rm etc.}
\ee
with the recursion relations,
\be \l{rec}
{{dF_{i,n}} \o {dy}}=F_{i-1,n}.
\ee
Thus we have arrived at the final form of the matrix $F$, which is as follows:
\be \l{fin}
F=\pmatrix{0&\cdots&0&0&g_0\cr
           0&\cdots&0&g_0&g_1\cr
           0&\cdots&g_0&g_1&g_2\cr
           \vdots&\vdots&\vdots&\vdots&\vdots\cr
           g_0&\cdots&g_{n-2}&g_{n-1}&g_n
           },
\ee
where each $g_i$ is a polynomial of degree $i$ in $y$, and $g_i={{dg_{i+1}}\o
{dy}}$. All the correlations depend on the $n$ constants $c_1,\cdots ,c_n$,
which remain undetermined. We have checked (as the reader can check for
the single $n=2$ case) that inversion symmetry puts no further restrictions
on the constants $c_i$.
\Section{n-Point Correlation Functions}

The observation (\ref{der}) that the transformation properties of the members
of a Jordan cell are as if the are formal derivative of the top field in the
cell, allows one to determine the correlation functions of all the fields within
a single or different Jordan cells, once the correlation function of
the top fields are determined. As an example, from ordinary CFT, we know
that conformal symmetry completely determines the three point function up to
a constant. Let $\phi _\alpha$, $\phi _\beta$ and $\phi _\gamma$ be the
top fields of three Jordan cells $(\Delta _\alpha ,l)$, $(\Delta _\beta ,m)$ and
$(\Delta _\gamma ,n)$ respectively. Therefore we know
that,
\be \l{ppp}
<\phi _\alpha (x)\phi _\beta (y)\phi _\gamma (z)>={{A_{\alpha\beta\gamma}}
\o {{\parallel x-y\parallel} ^{\Delta _\alpha +\Delta _\beta -\Delta _\gamma}
{\parallel x-z\parallel} ^{\Delta _\alpha +\Delta _\gamma -\Delta _\beta}
{\parallel y-z\parallel} ^{\Delta _\beta +\Delta _\gamma -\Delta _\alpha}}},
\ee
where the constant $A_{\alpha\beta\gamma}$ in principle depends on the weights
$\Delta _\alpha ,\Delta _\beta $ and $\Delta _\gamma$.
Denoting the second field of the cell $(\Delta _\alpha ,l)$ by $\phi _{
\alpha 1}$,\footnote {For simplicity, we have denoted the top field by $\phi$
and the second field by $\phi _1$, instead of $\phi _1$ and $\phi _2$
respectively.}
we will readily find from (\ref{ppp}) that,
\bea \l{three}
<\phi _{\alpha 1}(x)\phi _\beta (y)\phi _\gamma (z)>&=&-d {{\partial}\o{
\partial \Delta _\alpha}}<\phi _{\alpha }(x)\phi _\beta (y)\phi _\gamma (z)>
\nn\\
&=&
{
{A'_{\alpha\beta\gamma}} \o
{
{\parallel x-y\parallel} ^{\Delta _\alpha +\Delta _\beta -\Delta _\gamma}
{\parallel x-z\parallel} ^{\Delta _\alpha +\Delta _\gamma -\Delta _\beta}
{\parallel y-z\parallel} ^{\Delta _\beta +\Delta _\gamma -\Delta _\alpha}
}
}\nn\\&+&
{
{dA_{\alpha\beta\gamma}}
\o {
{\parallel x-y\parallel} ^{\Delta _\alpha +\Delta _\beta -\Delta _\gamma}
{\parallel x-z\parallel} ^{\Delta _\alpha +\Delta _\gamma -\Delta _\beta}
{\parallel y-z\parallel} ^{\Delta _\beta +\Delta _\gamma -\Delta _\alpha}
}}\ln \big ({{\parallel y-z\parallel }\o{(\parallel x-y\parallel )(\parallel
x-z\parallel )}}\big ),\nn\\
&&
\eea
where $A'_{\alpha\beta\gamma}=-d{{\partial} \o {\partial \Delta _\alpha}}
A_{\alpha\beta\gamma}$ is a new undetermined constant. For the correlation
functions
of fields within a single cell, one should then take the limit $\beta ,\gamma
\rightarrow \alpha$ in the above formula. It's not difficult to check that this
formula satisfies all the requirements demanded by conformal symmetry.
No need to say one can continue this procedure of formal differentiation
to determine other correlation functions.
It may be asked why we have not applied the procedure of formal differentiation
 for
calculation of two point function, but have followed an independent route. The
point
is that, the two point function has the form
$<\phi _\alpha (x)\phi _\beta (y)>={{A\delta _{\Delta _\alpha ,\Delta _\beta }}
\o {{\parallel x-y\parallel }^{2\Delta _\alpha}}}$, which upon differentiation,
 yields
derivative of the Kronecker symbol or the delta function which is not
easy to handle.
\Section{The Operator Product Expansion}

In this section we calculate the operator product of the fields in two different
Jordan cell, $(\Delta _{\alpha},N_\alpha)$, and$(\Delta _\beta ,N_\beta)$.
We assume that the operator product of the two top-fields $\phi _\alpha \in
(\Delta _\alpha ,N_\alpha )$ and $\phi _\beta \in (\Delta _\beta ,N_\beta )$
can be expressed
in terms of the fields and their descendants in other Jordan cells,
\be \l{ope}
\phi _\alpha (x)\phi _\beta (y)=\sum _{\gamma}\sum _{n=1}^{N_{\gamma}}
C_{\alpha \beta ,n}^\gamma (\parallel x-y\parallel )\phi _{\gamma ,n} (y),
\ee
where the sum is over the Jordan cells of weight $(\Delta _\gamma , N_\gamma)$.
The field $\phi _{\gamma ,n}$ stand for the $n-$th element of a Jordan cell
together with it's descendants. We
now demand that both sides have the same behaviour under conformal
transformations,
i. e. we demand that (\ref{ope}) is equivalent to the following relation,
\be \l{opeprime}
\phi '_\alpha (x')\phi '_\beta (y')=\sum _{n ,\gamma}
C_{\alpha \beta ,n}^\gamma (\parallel x'-y'\parallel )\phi '_{\gamma ,n} (y').
\ee
Considering first the scale transformation, we obtain from (\ref{confgt}),
\be \l{opescale}
\ll ^{-\Delta _\alpha -\Delta _\beta}\phi _\alpha (x)\phi _\beta (y)=
\sum _{n ,\gamma , p\leq N_{\gamma}} C_{\alpha \beta ,n}^\gamma (\ll (\parallel
 x-y\parallel ))
\ll ^{{-\Delta _\gamma}}\Lambda _{np}\phi _{\gamma ,p} (y).
\ee
Redefining the function $C_{\alpha \beta ,n}^\gamma (x)$ as follows,
\be \l{red}
C_{\alpha \beta ,n}^\gamma (x)={{C_{\alpha \beta}^\gamma f_n (x)}\o
{{\parallel x-y \parallel}^{\Delta _\alpha +\Delta _\beta -\Delta _\gamma}}},
\ee
where $C_{\alpha \beta }^\gamma$'s are constant and comparing eqs. (\ref{ope})
with (\ref{opescale}) leads to the following equation for $f$'s,
\be \l{fgh}
f_n (\ll ^{-1}x)=\sum _{p=1}^{N_\gamma}f_p (x)\Lambda _{pn},
\ee
or in compact matrix form,
\be \l{com}
f(\ll ^{-1}x)=\Lambda ^Tf(x),
\ee
Equation (\ref{com}) is easily
solved. It is in fact a recursion relation for $f_n$'s due to the triangular
form of $\Lambda$. The general solution is,
\be \l{fsol}
f_i (x)=\sum _{k=0}^{N-i}{{\alpha _{i+k}(-d\ln x)^k}\o {k}},
\ee
where the constants $\alpha _1, \cdots ,\alpha _N $ are free. For $N=3$ for
example, we have,
\bea \l{fn3}
f_3(x)&=&\alpha _3\nn\\
f_2(x)&=&\alpha _3(-d\ln x)+\alpha _2\nn\\
f_1(x)&=&{1 \o 2}\alpha _3(-d\ln x)^2+\alpha _2(-d\ln x)+\alpha _1.
\eea
It's interesting to note that the following relation holds among these
functions,
\be \l{fdif}
f_k(x)={{\p }\o {\p (-d\ln x)}}f_{k-1}(x).
\ee
\Section{Conjugation Properties}

In this section, we will restrict ourselves to two dmensions and derive the
hermitian conjugate of the fields in a Jordan cell. Our main rresult is encoded
 in the
following formula,
\be \l{ad}
{\hat \Phi}^\dagger (z)={\hat \Phi}^t ({{1}\o{\bar z}})\bar z
^{-2(\Delta -{J^t})}
\ee
where ${\hat \Phi}$ is an $N$ dimensional multiplet of operators, the
superscript
$t$, denotes transpose and the $\bar z$ dependence of the fields have been
supressed. As an example, when $N=2$ and ${\hat \Phi}=
\left (\begin{array}{c} {\hat \phi}\\ {\hat \psi}
\end{array} \right )$, we have,
\bea
{\hat \phi}^* (z)&=&{\bar z}^{-2\Delta}{\hat \phi}({{1}\o{\bar z}}),\nn\\
{\hat \psi}^* (z)&=&{\bar z}^{-2\Delta}({\hat \psi}({{1}\o{\bar z}})
+2\ln {\bar z}{\hat \phi}({{1}\o{\bar z}})).
\eea
As a justification for the validity of (\ref{ad}), consider the transformation
$z\rightarrow {{1}\o{\bar z}}$, with $\parallel z\parallel >
\parallel w\parallel$ and denote $<\phi _i(z)\phi _j(w)>$ by
$<\Phi (z)\Phi^t(w)>_{ij}$, then,
\be \l{add}
<\Phi (z)\Phi^t(w)>=<0\vert {\hat \Phi}(z){\hat \Phi}^t(w)\vert 0>=
<0\vert R{{\hat \Phi}^{\dagger t}}(w){\hat \Phi}^t(z)\vert 0>^{cc}
\ee
where cc denotes complex conjugation and $\vert 0>$ denotes the vaccum.
The right hand side of (\ref{add}) is then equal to,
\bea
<0\vert{\bar w}^{-2(\Delta -J)}{\hat \Phi}({{1}\o{\bar w}}){\hat \Phi}^t({{1}\o
{\bar z}}){\bar z}^{-2(\Delta -{J^t})}\vert 0>^{cc}&=&
w^{-2(\Delta -J)}<0\vert {\hat \Phi}({{1}\o{w}}){\hat \Phi}^t({1\o z})\vert 0>
z^{-2(\Delta -{J^t})}\nn \\
&=&w^{-2(\Delta -J)}<\Phi({{1}\o{w}}){\Phi}^t({1\o z})>
z^{-2(\Delta -{J^t})}\nn \\
&=&<\Phi '(w)\Phi '^t(z)>,
\eea
which is the desired equality.\newline
\noindent{\bf{Acknowledgement}}\newline
We would like to thank A. Aghamohammadi for valuable discussions.

\end{document}